\documentclass{aa}

\usepackage{txfonts}
\usepackage{graphicx}
\usepackage{rotating}
\usepackage{natbib}
\usepackage{lscape}
\usepackage{multirow}

\bibpunct{(}{)}{;}{a}{}{,} %to follow the A&A style
\newcommand{\kms}{km\,s$^{-1}$}

\newcommand{\lsun}{L$_{\odot}$}
\newcommand{\msun}{M$_{\odot}$}

\newcommand{\msunyr}{M$_{\odot}$\,yr$^{-1}$}

\begin{document}

\title{Towards a better classification of unclear eruptive variables:
the cases of V2492 Cyg, V350 Cep, and ASASSN-15qi}

\author{R. Jurdana-\v{S}epi\'{c}\inst{1},
        U. Munari\inst{2},
        S. Antoniucci\inst{3},
        T. Giannini\inst{3},
%        G. Li Causi\inst{3,4},
        D. Lorenzetti\inst{3}
                }

\institute{ Physics Department, University of Rijeka, Radmile Matej\v{c}i\'{c}, 51000 Rijeka, Croatia \and
            INAF - Osservatorio Astronomico di Padova - via dell’Osservatorio 8, Asiago (VI) 36012, Italy \and
            INAF - Osservatorio Astronomico di Roma - Via Frascati, 33 - Monte Porzio Catone 00078, Italy \and
%            INAF - Istituto di Astrofisica e Planetologia Spaziali - Via Fosso del Cavaliere, 100 - Roma 00133, Italy
                        } 

\offprints{Rajka Jurdana-\v{S}epi\'{c}, \email{jurdana@phy.uniri.hr}}
\date{Received date / Accepted date}
\titlerunning{Towards a better classification of unclear eruptive variables}
\authorrunning{Jurdana-\v{S}epi\'{c} et al.}

\abstract
{Eruptive variables are young stars that show episodic variations of brightness: EXors/FUors variations are commonly associated with enhanced accretion outbursts occurring at intermittent cadence of months/years (EXors) and decades/centuries (FUors). Variations that
can be ascribed to a variable extinction along their line of sight are instead classified as UXors.}
{We aim at investigating the long-term photometric behaviour of three sources classified as eruptive variables. We present data from the archival plates of the Asiago Observatory relative to
the fields where the targets are located. For the sake of completeness we have also analysed the Harvard plates of the same regions that cover a much longer historical period, albeit at a lower sensitivity, however we are only able to provide upper limits.}
{A total of 273 Asiago plates were investigated, providing a total of more than 200 magnitudes for the three stars, which cover 
a period of about 34 yr between 1958 and 1991. We have compared our data with more recently collected literature data.} 
{Our plates analysis of V2492 Cyg provides historical upper limits that seem not to be compatible with the level of the activity
monitored during the last decade. Therefore,  recently observed accretion phenomena could be associated with the outbursting episodes, more than repetitive obscuration. While a pure extinction does not seem the only mechanism responsible for the ASASSN-15qi fluctuations, it
can account quite reasonably for the recent V350 Cep variations.}
{}


\maketitle

\section{Introduction}

The last phases of the matter accretion onto young stellar objects (YSOs) occur through intermittent bursts of the mass-accretion rate that consequently provoke a remarkable increase of brightness, mainly at optical and near-infrared (NIR) frequencies.
A detailed description of the current knowledge of the accretion process onto young stars has recently been given by 
Hartmann, Herczeg, \& Calvet (2016). According to a general consensus these eruptive events are classified into two major classes, FUors and EXors, which, at the moment, globally amount to only a few tens of objects (see also the review by Audard et al. 2014).
FUors (Hartmann \& Kenyon 1985) are characterized by bursts of long duration (tens of years) with accretion rates of the order of 10$^{-4}$-10$^{-5}$ M$_{\odot}$~yr$^{-1}$ and spectra dominated by absorption lines. 
EXors (Herbig 1989, Lorenzetti et al. 2012) show shorter outbursts (between months and one year) with a recurrence time of years, have accretion rates of the order of 10$^{-6}$-10$^{-7}$ M$_{\odot}$~yr$^{-1}$, and are characterized by emission line spectra (e.g. Herbig 2008, Lorenzetti et al. 2009, K\'{o}sp\'{a}l et al. 2011, Sicilia-Aguilar et al. 2012, Antoniucci et al. 2013, 2014, Giannini et al. 2016).
Bursts of both classes are thought to be triggered by instabilities originating in the disk itself (Zhu et al. 2009, D'Angelo \& Spruit 2010, 2012), in particular in the inner connection region between star and disk. To have a more complete account for the scenario of the young variables, UXors objects must
also be considered (Grinin 1988), whose brightness variations are related to orbiting dust structures that move along the line of sight.

Observational activity is often oriented to obtain a quick classification by considering the mentioned
properties as overstrict, whereas it is becoming increasingly clear that a large variety of different cases exists, 
for which it is difficult to discriminate between the three classes (see e.g. the case of V1647 Ori, whose variability is loosely defined - Aspin 2011, K\'{o}sp\'{a}l et al. 2011;
%V2493 Cyg - Miller et al 2011, Semkov \& Peneva 2010, K\'{o}sp\'{a}l et al. 2011; 
and, more recently, that of V346 Nor - Kraus et al. 2016, K\'{o}sp\'{a}l et al. 2017). Sometimes even the origin of the observed variability is a matter of debate (see e.g. the case of GM Cep - Sicilia-Aguilar et al. 2008; Xiao et al. 2010). These dubious circumstances arise from observations that often rely on a single (or a few) event(s) and not on a long-lasting photometric and/or spectroscopic monitoring. Remarkably, a single outburst does not allow us to ascertain if the object will remain or not in its higher state; analogously, from a single fading, we cannot determine whether we are looking at a typically bright object subject to repetitive obscuration or, conversely, at a quiescent object that undergoes recurrent outbursts.

One method to ameliorate the classification process involves investigating (whenever it is possible) the past
history of the objects. A long-lasting (up to half century or more) monitoring can be obtained by digging into the plate archives, which allow us to build up light curves in the optical bands, where the eruptive variables present their largest
fluctuations. This method has proven to be very efficient for studying the eruptive variables, as recently demonstrated for the FUor V960 Mon (Jurdana--\v{S}epi\'{c} \& Munari, 2016) and for the EXors GM Cep (Xiao et al. 2010) and V1118 Ori (Jurdana--\v{S}epi\'{c} et al. 2017).

In the following, we examine deep $BVRI$ plates taken between 1958
and 1993 with the Asiago Schmidt telescopes, of three young variables classified as eruptive objects, namely V2492
Cyg, V350 Cep, and ASASSN-15qi, which for different reasons deserve a more in-depth analysis in order to ascertain their nature.  We aim at improving the characterization of
their quiescent phase, which will represent a useful reference for studies
analysing future outbursts and the physical changes induced by these
events.  

To expand our search to a more distant past (up to $\sim$1890),
we have also examined all the deepest plates covering our three program stars
that we have been able to locate in the plate stack of the Harvard College 
Observatory (HCO). None of them, unfortunately, turned out to be as
deep as the Asiago plates, their limiting recorded brightness being typically
several magnitudes brighter.

The paper is organised as follows: our sample is presented in Sect. 2; the adopted method and the obtained $BVRI$ photometry are presented in Sects. 3 and 4. Section 5 gives the analysis and discussion of the obtained results, while our concluding remarks are given in Sect. 6.

\section{The investigated sample} 

\subsection{V2492 Cyg} 
The object V2492 Cyg ($\alpha_{2000}$ = 20$^{h}$51$^{m}$26.23$^{s}$,
$\delta_{2000}$ +44$^{\circ}$05$^{\prime}$23.9$^{\prime \prime}$) is located in the North American-Pelican Nebula at an estimated distance of 550 pc (Bally \& Reipurth 2003).
Its first outburst on 2010 August 23 was discovered by Itagaki \& Yamaoka (2010) and the brightness increase of $\sim$5 mag lasted about 10 months. Covey et al. (2011) independently discovered the V2492 Cyg outburst: their optical and near-IR spectroscopy revealed a rich emission-line spectrum that allowed them to derive a mass-accretion rate of 2.5 10$^{-7}$ \msunyr. Blue-shifted absorption at several hundred  \kms and the presence of typical tracers of shocked gas (such as H$_2$, [SII] and [FeII]) both indicate a substantial amount of outflowing matter. The same spectroscopic evidence was confirmed by Aspin (2011), who also considered archival data (2009-2010) to investigate the pre-outburst nature of V2492 Cyg. His conclusions favour an EXor classification for V2492 Cyg, although the source appears to be significantly younger than other members typical of the class.
After its fading, the source underwent additional episodes of brightening followed by dimming events; all these phases
were extensively sampled by Hillenbrand et al. (2013), who detected neutral and singly ionized atomic species likely formed in an accretion flow, but also identified a behaviour attributable to rotating circumstellar disk material that causes the semi-periodic dimming. Therefore they concluded that V2492 Cyg simultaneously displays accretion-driven and extinction-driven fluctuations.
K\'{o}sp\'{a}l et al. (2013), observing the source between 0.55 $\mu$m and 160 $\mu$m, regard V2492 Cyg as an 
UXor-type system. The most recent (Nov. 2016 - Mar. 2017) peak of brightness, at a level never reached before, has been announced by Ibryamov \& Semkov (2017) and spectroscopically studied at high resolution by Munari et al. (2017) and Giannini et al. (2017), who confirm the great deal of emission lines as presented by Aspin (2011), but with absorptions that are now far stronger.\\

\subsection{V350 Cep} 
Herbig (2008) provides a summary of the past history
of the late-type (M2) pre-main sequence star V350 Cep ($\alpha_{2000}$ = 21$^{h}$43$^{m}$00.00$^{s}$,
$\delta_{2000}$ +66$^{\circ}$11$^{\prime}$28.0$^{\prime \prime}$) in NGC 7129, a region of active star formation (Dahm \& Hillenbrand 2015 and references therein) located at 1150 pc (Strai\u{z}ys et al. 2014). The target was undetected (i.e. $B>$21 mag) on the 1954 Palomar plates, but a brightness (about $B$= 17.5) was found in the middle of the 1970s. From 1978 until 2004, V350 Cep remained at the same level of brightness ($B\sim$17) as shown by Pogosyants (1991), and Semkov (2004, and references therein). 
As a consequence, all the following spectroscopic investigations were obtained at a high level of brightness, and, substantially, all confirm the classical T Tauri (CTTS) nature of V350 Cep. 
Herbig (2008) also examined the high-resolution spectrum discussing important dynamical details, suggesting that V350 cannot be classified
as an EXor source. Ibryamov et al. (2014) provide the result of their $UBVRI$ photometry in the period 2004-2014, during which the star maintained its maximum brightness (for a complete view of the historical light curve, see their Fig.~1). The V-band light curve retrievable from the ASASSN  \footnote{All-Sky Automated Survey for Supernovae (https://asas-sn.osu.edu/)} database
confirms that V350 Cep is even now at the same high level of brightness.
The simultaneous presence of an emission line spectrum and the apparent lack of a repetitive activity represent contrasting evidences toward a certain classification of V350 Cep.
Further ambiguities stem from the recent observations by Semkov et al. (2017), who registered a deep fading ($\Delta B$ = 2.16, $\Delta V$ = 1.77) in the period between March and May 2016, followed by a quick restoration (occurring in the second half of 2016) of its maximum brightness.
They suggest that such an event is compatible with an obscuration from clumps orbiting the star or with a decrease in the accretion rate, or with a combination of both mechanisms.\\

\subsection{ASASSN-15qi} 
ASASSN-15qi ($\alpha_{2000}$ = 20$^{h}$56$^{m}$08.82$^{s}$,
$\delta_{2000}$ +51$^{\circ}$31$^{\prime}$04.1$^{\prime \prime}$) recently (2015 Oct 2nd/3rd)
underwent a sudden brightening ($\Delta$V = 3.5 mag) in less than one day.
Strong P-Cyg profiles with red emission wings traced a very fast wind with a velocity up to 1000 
\kms\  that faded while the central source returned to quiescence (Maehara et al. 2015, Herczeg et al. 2016).
The distance estimate for ASASSN-15qi is around 3.24 kpc. 
Assuming the values of the pre-outburst photometry, Hillenbrand et al. (2015) derived an A$_V$ 
of about 5 mag  and a luminosity of $\sim$125 \lsun~. Consequently, they also derived 
a progenitor source roughly 2.5-3.0 \msun, and concluded the source cannot be easily classified into the
FUor and EXors classes according to our current knowledge of the defining criteria. 
Connelley et al. (2015) obtained an early near-IR spectrum showing a strong veiling, a number of emission lines (with strong P-Cyg), and CO bandheads in emission. They also revealed a very faint nebulosity seen only in the $J$ band), likely due to a 
reflection nebula. A thorough collection of UV, optical, near- and mid-IR and sub-mm observations, 
together with archival photometry (mainly obtained in 2000 and 2015), was presented by Herczeg el al.
(2016). They were able to build-up reliable Spectral Energy Distributions (SEDs) of ASASSN-15qi during quiescence, outburst, peak, decay, and 
quiescence again. They also provide an accurate description of the dynamical events during the different 
phases. Nevertheless, the observations cannot easily be explained in the framework of FUor and EXor outbursts, 
and, in principle, even a rapid decrease in the extinction could be responsible for the increased brightness.
The V-band light curve (2015-2017) from the ASASSN database provides significant upper limits (V $<$ 17 mag) confirming that ASASSN-15qi is currently in a quiescence period\

\section{Data acquisition} 

\subsection{Asiago archive plates}

Two Schmidt telescopes were operated with photographic plates at the
Astronomical Observatory in Asiago.  The smaller one (SP, 40/50 cm, 100 cm
focal length) observed between 1958 and 1992, with photographic films
covering a circular area 5$^\circ$ in diameter.  The larger telescope
(SG, 67/92 cm, 208 cm focal length) exposed square glass photographic plates
imaging a 5$^\circ\times$5$^\circ$ portion of the sky.  It operated with 
photographic plates between 1965 and 1998, after which large format CCDs were
used as detectors.  Nearly all plates from both telescopes
go very deep, $B$$\sim$18.5 and $B$$\sim$17.8 mag being the typical limiting
magnitude for blue sensitive plates exposed with the SG and SP
telescopes, respectively.

The Asiago plates were usually exposed as 103a-O + GG13, 103a-E + RG1, Tri-X
+ GG14, and IN + RG5 combinations of Kodak plates and Schott astronomical
filters, matching the passband of Johnson-Cousins $B$, $R_{\rm C}$ and
$I_{\rm C}$ photometric system (Moro \& Munari 2000).  A large number of
plates were exposed as unfiltered 103a-O, thus nominally covering both the
Johnson $B$ and $U$ bands thanks to the high ultraviolet transparency of the
UBK-7 corrector plates at both Schmidt telescopes.  For low temperature
and/or reddened objects, especially if they were observed at large airmass,
the amount of proper $U$-band photons collected by an unfiltered 103a-O plate 
is however minimal compared to those arriving through the $B$-band portion 
of the interval of sensitivity of 103a-O emulsion.  In such conditions
(and provided that the selected comparison stars are themselves of low
temperature and/or high reddening), 103a-O + GG13 pairs and unfiltered
103a-O plates are almost equally well replicating the standard Johnson $B$
band (Munari \& Dallaporta 2014).

The Asiago Schmidt plates cover an uninterrupted interval of 40 years,
and are accurately preserved in a controlled environment. We note that
their time spans overlap the so-called {\it Menzel Gap} when for about
ten years the acquisition of plates with Harvard telescopes distributed 
around the globe was greatly reduced. A total of 90, 68, and 115 Asiago plates
were found to image, respectively, the three program stars V2492 Cyg, V350
Cep and ASASSN-15qi. The resulting magnitudes are listed in Tables 1, 2, 
and 3 (electronic only).

\subsection{Harvard plates}

In order to expand our search for old starbursts of our program stars as far back as possible into the past, we turned to the Harvard College Observatory
(HCO) plate stack in Cambridge, Massachusetts, home of one of the largest
and best preserved archives of astronomical photographic plates taken with
many different astrographs located at various sites distributed in both
hemispheres beginning around 1880.  The limited focal length and aperture of
these astrographs result in detection limits which are typically some
magnitudes brighter than those of the far deeper Asiago plates.

V350 Cep by lucky coincidence happens to be located in a region of the sky
where all HCO plates have been digitized and measured as part of the DASCH
project (Grindlay et al.  2012), that we were allowed to access prior to
publication (E. Los, 2017 private communication).  The other two targets,
V2492 Cyg and ASASSN-15qi, are in regions of the sky not covered by DASCH,
so we personally went to the HCO plate stack, located and retrieved the
relevant plates, and inspected them at the microscope against local
photometric sequences extracted from the all-sky APASS photometric survey
(Henden \& Munari 2014).  The process is tedious and therefore we decided to
limit the inspection to HCO plates from instruments that were expected
to pass a limiting magnitude of 13 in B band (almost all of the old HCO
plates are blue sensitive with a response comparable to that of modern B
band).

About 535 plates for V2492 Cyg and 162 for ASASSN-15qi were selected for
visual inspection.  About 198 for V2492 Cyg and 40 for ASASSN-15qi were
found to be inadequate for the task for a variety of reasons (emulsion defects,
fogging, poor guiding/seeing, plate missing or imaging an incorrect field, etc.),
leaving 337 suitable plates for V2492 Cyg and 122 for ASASSN-15qi.  In none
of them were the program stars detected.  The same holds true for V350 Cep,
which is undetected on all plates digitized by DASCH.  A random selection of
the latter was inspected at the microscope to compare our estimates with
DASCH limiting magnitudes, and similar results were found (incidentally, the
photometric calibration of DASCH scans is done based on the APASS survey too).  
Figure~\ref{Harvard:fig} presents the upper limits from
HCO plates to the brightness of our program stars in graphical form.  
 
\subsection{Brightness measurement}

To derive the brightness of our targets, we compared them at a high quality
Zeiss microscope against a local photometric sequence established around
each target.  Such a sequence, composed of stars of roughly the same colour
as the variable, and widely distributed in magnitude so as to cover both
quiescence and outburst states, was extracted primarily from the APASS
$B$$V$$g'$$r'$$i'$ all-sky survey (Henden et al.  2012, Henden \& Munari
2014), with porting to Landolt $R_{\rm C}$ and $I_{\rm C}$ bands following
Munari (2012) and Munari et al.  (2014) prescriptions.  To evaluate the
measurement errors a number of plates were re-estimated after several days
when all memories had vanished from the observer(s).  The typical error is
0.1 mag, comparable to that intrinsic to the photographic plate itself so
that the observer adds little to it.  For plates not recording the variable
star due to it being too faint, we noted the magnitude of the faintest but still
clearly visible star of the comparison sequence.  The latter was composed
by field stars closely distributed around the variable and typically by 0.35
mag in brightness.

\begin{figure*}
\centering
\includegraphics[width=15cm, trim = 0cm .5cm 0cm 1cm, clip]{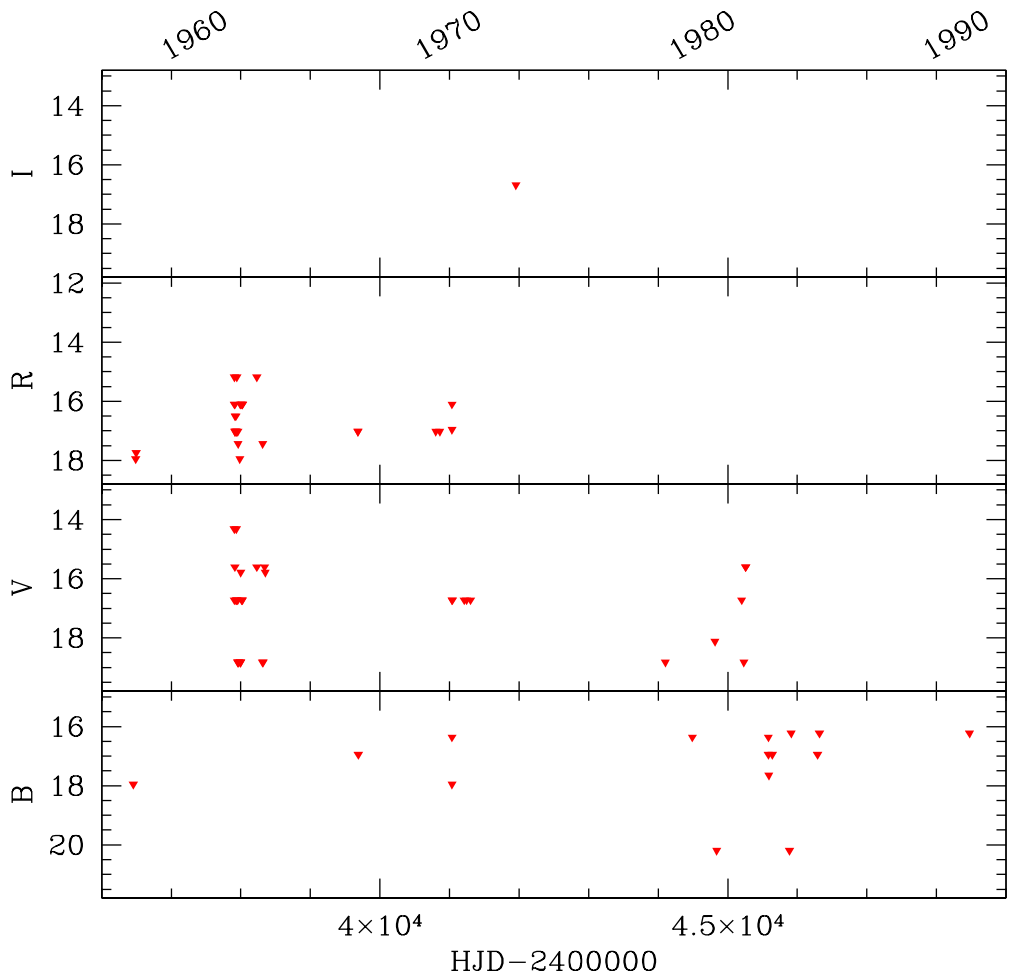}
\caption{\label{V2492_light:fig} $BVRI$ magnitude upper limits (solid red triangles) for V2492 Cyg. 
%The recent (2010-2017) activity is depicted as black symbols between the two vertical dashed lines. Such an activity is repeatedly superimposed onto our upper limits hypothesizing that it obeys a period of about 220 days (see text).
}
\end{figure*}

\begin{figure*}
\centering
\includegraphics[width=15cm, trim = 0cm .5cm 0cm 1cm, clip]{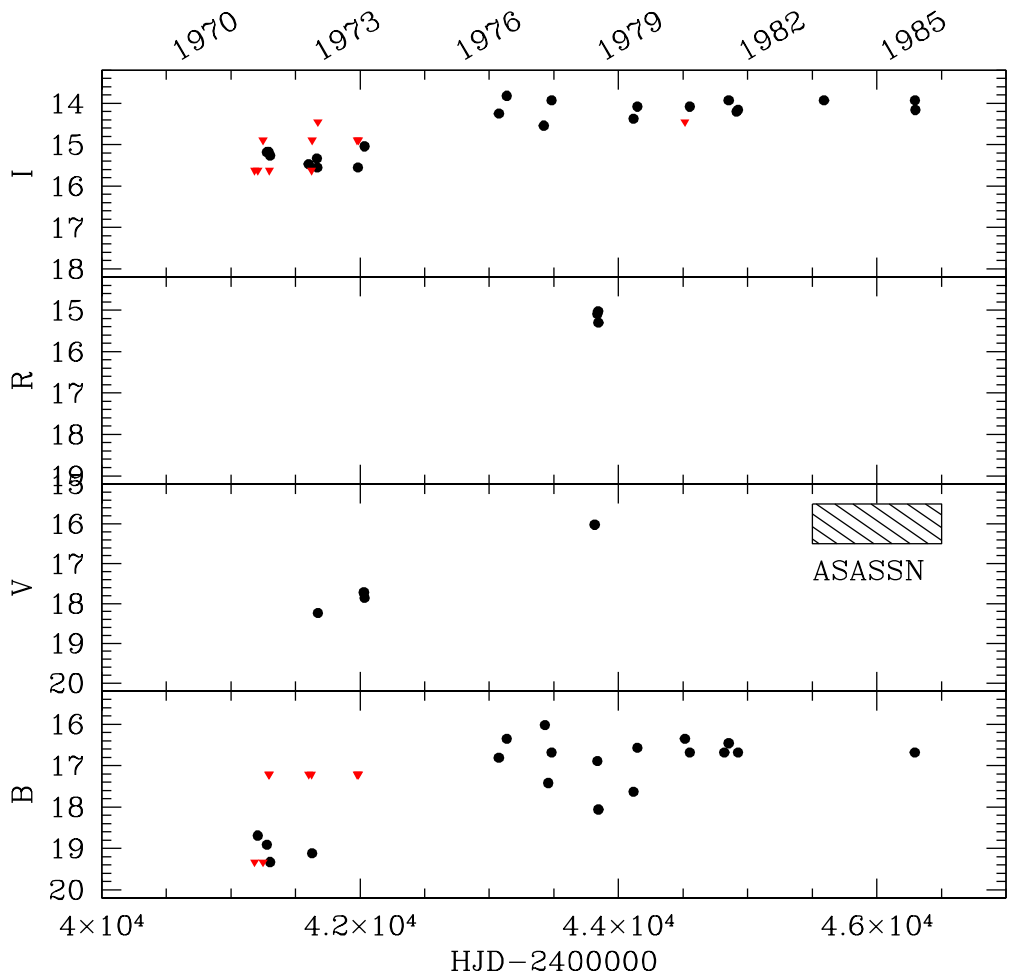}
\caption{\label{V350_light:fig} $BVRI$ light curves of V350 Cep. Upper limits are given as solid red triangles. The hatched box on the right-hand side of the V-band panel indicates the spread of magnitudes obtained in the period 2014-2017 by the survey ASASSN (see text).
}
\end{figure*}

\begin{figure*}
\centering
\includegraphics[width=15cm, trim = 0cm .5cm 0cm 1cm, clip]{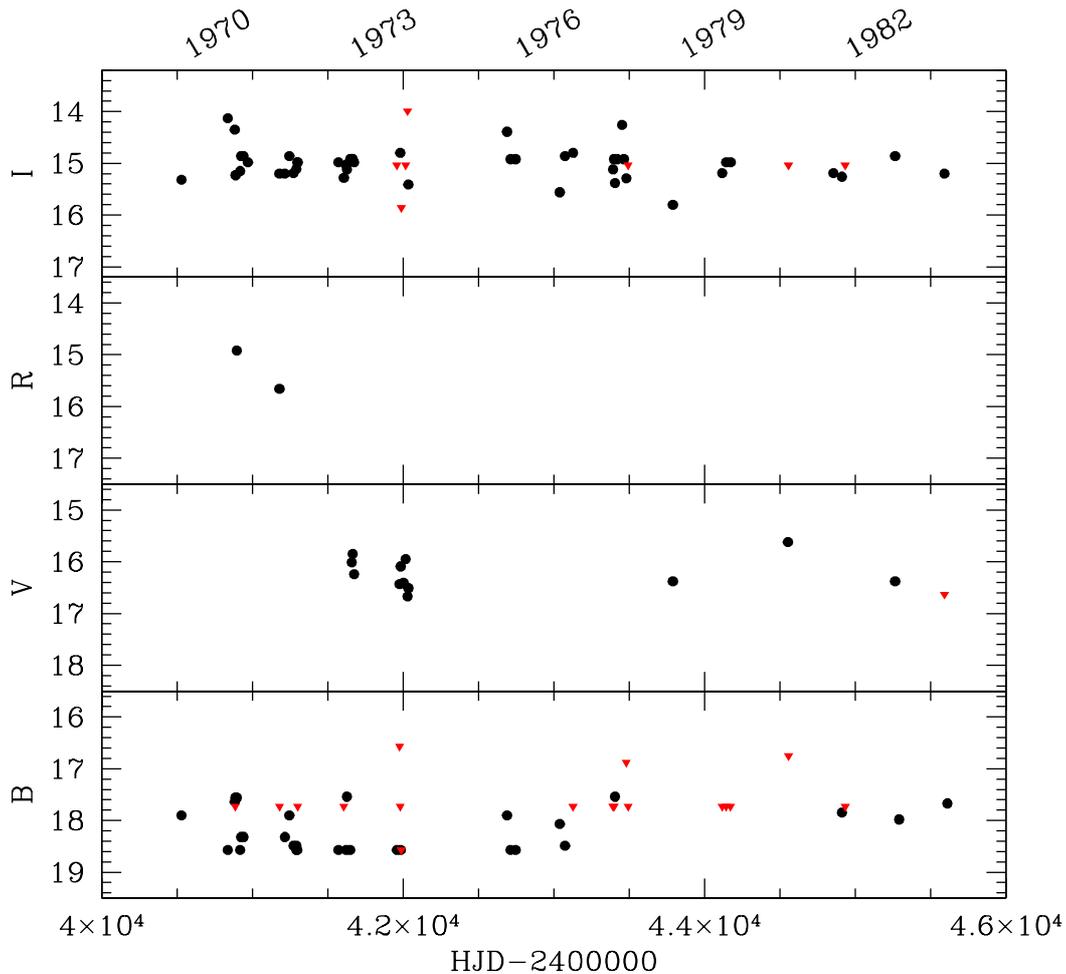}
\caption{\label{ASASSN_light:fig} $BVRI$ light curves of ASASSN 15qi. Upper limits are given as solid red triangles. The horizontal segment in the right side of the V-band panel indicates the level of upper limit magnitudes obtained in the period 2015-2017 by the survey ASASSN (see text). 
}
\end{figure*}

\begin{figure*}
\centering
\includegraphics[width=13cm]{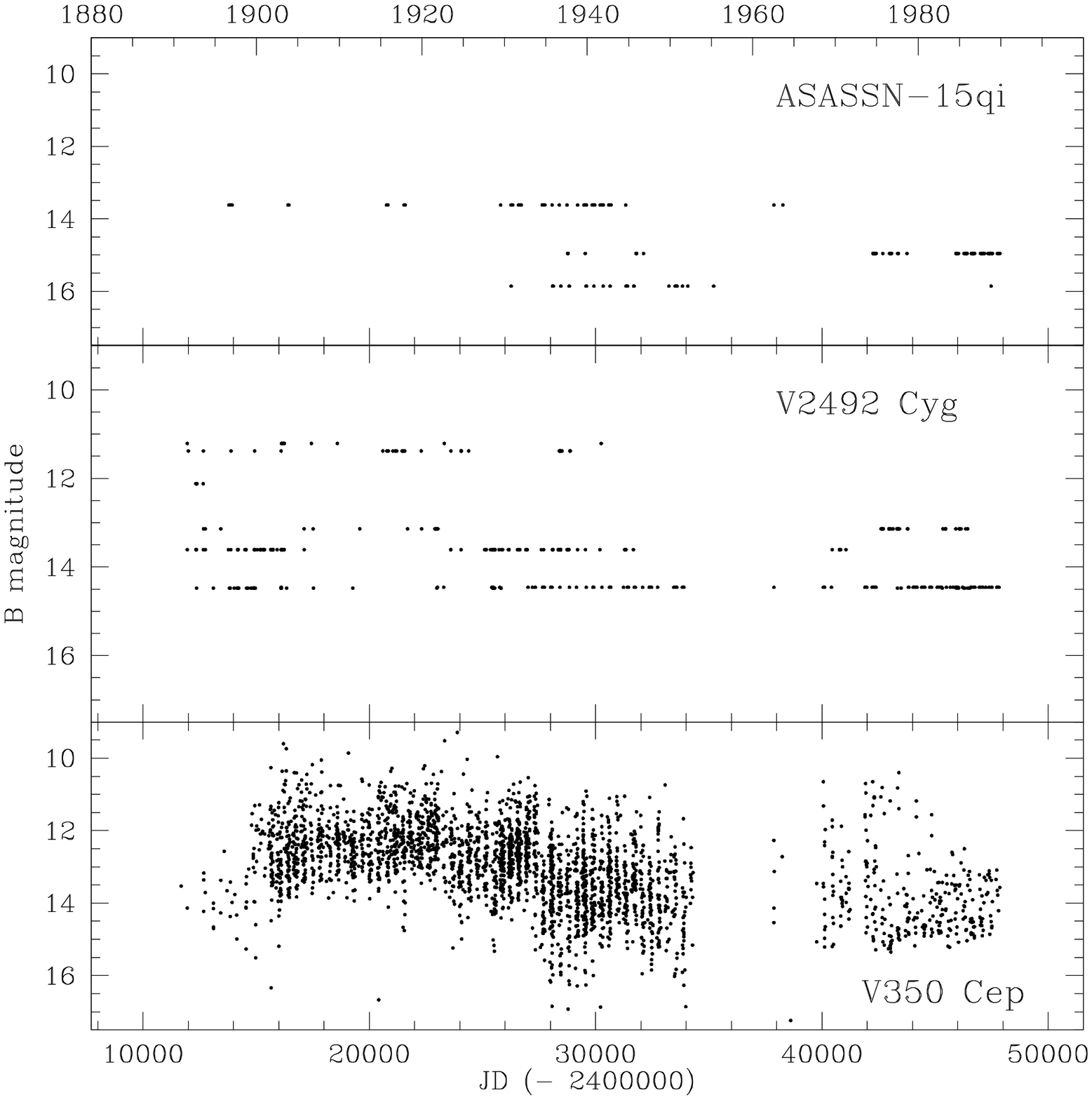}
\caption{\label{Harvard:fig} Limiting magnitude for the Harvard plates inspected
in the search for past bright episodes of the program stars.  For V2492 Cyg and
ASASSN-15qi, we have plotted  the B-band magnitude of the faintest APASS star
visible at the microscope in the immediate surroundings of the targets.  For
V350 Cep we plot the limiting magnitude as derived by DASCH as part of the
calibrations of the plate scans.
}
\end{figure*}

\section{Historical light curves} 

Because of the limits imposed by the Harvard plate, the analysis presented hereinafter will be based  on the Asiago plates only.
The $BVRI$ light curves of V2492 Cyg, V350 Cep, and ASASSN 15qi corresponding to their plate photometry (Tables \ref{photometry_V2492:tab}, \ref{photometry_V350:tab} and \ref{photometry_ASASSN:tab}) are given in Figures~\ref{V2492_light:fig}, \ref{V350_light:fig}, and \ref{ASASSN_light:fig}, respectively. In Table~\ref{ranges:tab} some statistics are provided for the 
source  ASASSN-15qi, the only one for which we can infer, for the first time, meaningful averaged values for the quiescence level.
Determining this level is fundamental to have a solid reference for accurately computing physical changes once the outburst values are obtained. In each band, we list the number of observations, the median value (basically the quiescence magnitude) together with the standard deviation of the data points, and the peak brightness. We note how this source presents, in quiescence, a level of modest variability quantified by considering the standard deviation of the measurements (0.3-0.4 mag).

%\setcounter{table}{0}
%%%%%%%%%%%%%%%%%%%%%%%%%%%%%%%%%%%%%%%%%%%%%%%%%%%%%%%%%%%%%%%%%%%%%%%%%%%%%%%%%%%%%%%%%%%%%%%%%%% 
\begin{table*}
\caption{Plate photometry of V2492 Cyg. Columns provide: date and central UT of any exposure, exposure time, plate emulsion, the adopted filter, the telescope, the plate number, and the magnitude derived for the source (see text for further details).  \label{photometry_V2492:tab}} 
\begin{tiny}
\begin{tabular}{c|ccr|llcr|l}
\hline
\hline
\multicolumn{1}{c}{HJD}&
\multicolumn{1}{c}{Date}&
\multicolumn{1}{c}{UT}&
\multicolumn{1}{c}{Expt}&
\multicolumn{1}{c}{Emulsion}&
\multicolumn{1}{c}{Filter}&
\multicolumn{1}{c}{Tel}&
\multicolumn{1}{c}{Plate}&
\multicolumn{1}{c}{mag}\\
\hline
2436454.42875 & 1958-09-07      & 22:13 & 30    & 103a-O        &....    & SP  & 00031  & B$>$ 17.95  \\ 
2436488.31134 & 1958-10-11      & 19:25 & 60    & 103a-E        &RG 1    & SP  & 00049  & R$>$ 17.95  \\ 
2436494.29375 & 1958-10-17      & 19:00 & 40    & 103 a-E       &RG 1    & SP  & 00052  & R$>$ 17.74  \\ 
2437905.46837 & 1962-08-28      & 23:10 & 15    & PANROY        &....    & SP  & 03028  & V$>$ 14.32  \\ 
2437907.44614 & 1962-08-30      & 22:38 & 15    & PANROY        &....    & SP  & 03037  & V$>$ 16.72  \\ 
2437907.46420 & 1962-08-30      & 23:04 & 15    & 103 a-E       &RG 1    & SP  & 03038   & R$>$ 15.18  \\ 
2437908.48295 & 1962-08-31      & 23:31 & 40    & 103 a-E       &RG 1    & SP  & 03045   & R$>$ 16.10  \\ 
2437913.51001 & 1962-09-05      & 24:10 & 15    & PANROY        &....    & SP  & 03055  & V$>$ 15.60  \\ 
2437913.55098 & 1962-09-05      & 25:09 & 30    & 103 a-E       &RG 1    & SP  & 03056   & R$>$ 16.10  \\ 
2437915.47180 & 1962-09-07      & 23:15 & 30    & 103 a-E       &RG 1    & SP  & 03060   & R$>$ 17.02  \\ 
2437916.44819 & 1962-09-08      & 22:41 & 40    & 103 a-E       &RG 1    & SP  & 03065   & R$>$ 17.02  \\ 
2437917.48776 & 1962-09-09      & 23:38 & 40    & 103 a-E       &RG 1    & SP  & 03075   & R$>$ 16.50  \\ 
2437926.33139 & 1962-09-18      & 19:53 & 30    & 103 a-E       &RG 1    & SP  & 03083   & R$>$ 17.02  \\ 
2437929.32509 & 1962-09-21      & 19:44 & 40    & 103 a-E       &RG 1    & SP  & 03095   & R$>$ 16.50  \\ 
2437930.37645 & 1962-09-22      & 20:58 & 15    & PANROY        &....    & SP  & 03101  & V$>$ 16.72  \\ 
2437933.46666 & 1962-09-25      & 23:08 & 40    & 103 a-E       &RG 1    & SP  & 03118   & R$>$ 17.02  \\ 
2437936.52909 & 1962-09-28      & 24:38 & 15    & PANROY        &....    & SP  & 03133  & V$>$ 14.32  \\ 
2437938.42557 & 1962-09-30      & 22:09 & 15    & PANROY        &....    & SP  & 03146  & V$>$ 16.72  \\ 
2437939.40401 & 1962-10-01      & 21:38 & 15    & PANROY        &....    & SP  & 03161  & V$>$ 16.72  \\ 
2437940.35468 & 1962-10-02      & 20:27 & 40    & 103 a-E       &RG 1    & SP  & 03170   & R$>$ 17.02  \\ 
2437942.37337 & 1962-10-04      & 20:54 & 40    & 103 a-E       &RG 1    & SP  & 03189   & R$>$ 15.18  \\ 
2437955.28264 & 1962-10-17      & 18:44 & 15    & PANROY        &....    & SP  & 03208  & V$>$ 16.72  \\ 
2437956.29024 & 1962-10-18      & 18:55 & 30    & 103 a-E       &RG 1    & SP  & 03213   & R$>$ 17.02  \\ 
2437957.28742 & 1962-10-19      & 18:51 & 20    & PANROY        &....    & SP  & 03219  & V$>$ 18.82  \\ 
2437960.28591 & 1962-10-22      & 18:49 & 40    & 103 a-E       &RG 1    & SP  & 03225   & R$>$ 17.43  \\ 
2437961.27892 & 1962-10-23      & 18:39 & 20    & PANROY        &....    & SP  & 03240  & V$>$ 18.82  \\ 
2437962.30805 & 1962-10-24      & 19:21 & 20    & PANROY        &....    & SP  & 03249  & V$>$ 18.82  \\ 
2437985.23194 & 1962-11-16      & 17:33 & 40    & 103a-E        &RG 1    & SP  & 03294  & R$>$ 17.95  \\ 
2437991.24829 & 1962-11-22      & 17:57 & 20    & PANROY        &....    & SP  & 03298  & V$>$ 18.82  \\ 
2437992.23296 & 1962-11-23      & 17:35 & 40    & 103 a-E       &RG 1    & SP  & 03311   & R$>$ 16.10  \\ 
2437995.21822 & 1962-11-26      & 17:14 & 20    & PANROY        &....    & SP  & 03337  & V$>$ 18.82  \\ 
2437998.21251 & 1962-11-29      & 17:06 & 15    & PANROY        &....    & SP  & 03347  & V$>$ 18.82  \\ 
2437999.23328 & 1962-11-30      & 17:36 & 40    & 103 a-E       &RG 1    & SP  & 03367   & R$>$ 16.10  \\ 
2438000.23601 & 1962-12-01      & 17:40 & 10    & PANROY        &....    & SP  & 03382  & V$>$ 15.78  \\ 
2438013.25477 & 1962-12-14      & 18:08 & 15    & PANROY        &....    & SP  & 03440  & V$>$ 16.72  \\ 
2438015.21369 & 1962-12-16      & 17:09 & 30    & 103 a-E       &RG 1    & SP  & 03447   & R$>$ 16.10  \\ 
2438017.21429 & 1962-12-18      & 17:10 & 30    & 103 a-E       &RG 1    & SP  & 03466   & R$>$ 16.10  \\ 
2438019.23502 & 1962-12-20      & 17:40 & 15    & PANROY        &....    & SP  & 03483  & V$>$ 16.72  \\ 
2438020.24677 & 1962-12-21      & 17:57 & 30    & 103 a-E       &RG 1    & SP  & 03493   & R$>$ 16.10  \\ 
2438021.21201 & 1962-12-22      & 17:07 & 15    & PANROY        &....    & SP  & 03507  & V$>$ 16.72  \\ 
2438024.21394 & 1962-12-25      & 17:10 & 30    & 103a-E        &RG 1    & SP  & 03525  & R$>$ 16.10  \\ 
2438229.49955 & 1963-07-18      & 23:56 & 10    & PANROY        &....    & SP  & 04082  & V$>$ 15.60  \\ 
2438230.50514 & 1963-07-19      & 24:04 & 30    & 103 a-E       &RG 1    & SP  & 04091   & R$>$ 15.18  \\ 
2438311.30172 & 1963-10-08      & 19:11 & 13    & PANROY        &....    & SP  & 04095  & V$>$ 18.82  \\ 
2438314.31482 & 1963-10-11      & 19:30 & 30    & 103 a-E       &RG 1    & SP  & 04108   & R$>$ 17.43  \\ 
2438321.30761 & 1963-10-18      & 19:20 & 22    & PANROY        &....    & SP  & 04150  & V$>$ 18.82  \\ 
2438342.26500 & 1963-11-08      & 18:20 & 15    & PANROY        &....    & SP  & 04229  & V$>$ 15.60  \\ 
2438351.24371 & 1963-11-17      & 17:50 & 15    & PANROY        &....    & SP  & 04238  & V$>$ 15.78  \\ 
2439681.53672 & 1967-07-09      & 24:50 & 60    & 103a-E        &RG 1    & SG  & 00741  & R$>$ 17.02  \\ 
2439685.47923 & 1967-07-13      & 23:27 & 90    & 103a-E        &RG 1    & SG  & 00754  & R$>$ 17.02  \\ 
2439688.53768 & 1967-07-16      & 24:51 & 30    & 0a-O          &        & SG  & 00768  & B$>$ 16.95  \\ 
2440802.49793 & 1970-08-03      & 23:53 & 60    & Ia-E          &RG 6450 & SG  & 03527  & R$>$ 17.02  \\ 
2440859.42767 & 1970-09-29      & 22:12 & 60    & 103a-E        &RG 6450 & SG  & 03654  & R$>$ 17.02  \\ 
2441035.62019 & 1971-03-24      & 26:57 & 15    & 103 a-O       &GG 13   & SP  & 08717   & B$>$ 16.36  \\ 
2441035.62575 & 1971-03-24      & 27:05 & 10    & 103a-O        &GG 13   & SG  & 04262  & B$>$ 17.95  \\ 
2441035.63616 & 1971-03-24      & 27:20 & 10    & 103 a-E       &RG 1    & SP  & 08718   & R$>$ 16.95  \\ 
2441035.63686 & 1971-03-24      & 27:21 & 10    & 103a-D        &GG 14   & SG  & 04263  & V$>$ 16.72  \\ 
2441035.65144 & 1971-03-24      & 27:42 & 20    & 103a-E        &RG 1    & SG  & 04264  & R$>$ 16.10  \\ 
2441040.64879 & 1971-03-29      & 27:38 & 15    & 103a-D        &GG 14   & SG  & 04281  & V$>$ 16.72  \\ 
2441213.40778 & 1971-09-18      & 21:43 & 20    & TRI X         &GG 14   & SP  & 09022  & V$>$ 16.72  \\ 
2441243.45344 & 1971-10-18      & 22:50 & 20    & TRI X         &GG 14   & SP  & 09122  & V$>$ 16.72  \\ 
2441303.29768 & 1971-12-17      & 19:10 & 20    & TRI X         &GG 14   & SP  & 09312  & V$>$ 16.72  \\ 
2441954.46936 & 1973-09-28      & 23:12 & 20    & IN Sen        &RG 5    & SG  & 06655  & I$>$ 16.69  \\ 
2444101.43147 & 1979-08-15      & 22:17 & 30    & 103a-D        &GG 14   & SG  & 10133  & V$>$ 18.82  \\ 
2444490.47666 & 1980-09-07      & 23:22 &  8    & 103 a-O       &GG 13   & SP  & 14349   & B$>$ 16.36  \\ 
2444814.50196 & 1981-07-28      & 23:59 & 30    & 103a-D        &GG 14   & SG  & 11050  & V$>$ 18.12  \\ 
2444841.49961 & 1981-08-24      & 23:55 & 30    & 103a-O        &GG 13   & SG  & 11084  & B$>$ 20.19  \\ 
2445199.48844 & 1982-08-17      & 23:39 & 15    & TRI X         &....    & SP  & 14906  & V$>$ 16.72  \\ 
2445231.39041 & 1982-09-18      & 21:18 & 30    & 103a-D        &GG 14   & SG  & 11682  & V$>$ 18.82  \\ 
2445258.38896 & 1982-10-15      & 21:17 & 15    & TRIX          &....    & SP  & 14982  & V$>$ 15.60  \\ 
2445259.39656 & 1982-10-16      & 21:28 & 15    & TRIX          &....    & SP  & 14994  & V$>$ 15.60  \\ 
2445582.38849 & 1983-09-04      & 21:15 & 15    & 103 a-O       &GG 13   & SP  & 15354   & B$>$ 16.95  \\ 
\hline
\end{tabular}
\end{tiny}
%\tablefoot{
%\tablefoottext{a}{A},\tablefoottext{b}{B},\tablefoottext{c}{C},\tablefoottext{d}{D}\\
%}
\end{table*}  
\setcounter{table}{0}
\begin{table*}
\caption{Continued.} 
\begin{tiny}
\begin{tabular}{c|ccr|llcr|l}
\hline
\hline
2445583.38015 & 1983-09-05      & 21:03 & 15    & 103 a-O       &GG 13   & SP  & 15361   & B$>$ 16.36  \\ 
2445590.41065 & 1983-09-12      & 21:47 & 10    & 103 a-O       &....    & SP  & 15380   & B$>$ 17.65  \\ 
2445641.38819 & 1983-11-02      & 21:17 & 12    & 103 a-O       &....    & SP  & 15480   & B$>$ 16.95  \\ 
2445886.49766 & 1984-07-04      & 23:54 & 30    & 103a-O        &GG 13   & SG  & 12519  & B$>$ 20.19  \\ 
2445912.46799 & 1984-07-30      & 23:10 & 10    & 103 a-O       &....    & SP  & 15930   & B$>$ 16.23  \\ 
2446290.50297 & 1985-08-12      & 24:00 & 12    & 103a-O        &....    & SP  & 16379  & B$>$ 16.95  \\ 
2446318.42526 & 1985-09-09      & 22:08 & 10    & 103a-O        &....    & SP  & 16461  & B$>$ 16.23  \\ 
2448476.44454 & 1991-08-07      & 22:36 & 10    & 103 a-O       &GG 13   & SP  & 18248  & B$>$ 16.23  \\ 
2448539.46904 & 1991-10-09      & 23:12 &  8    & 103 a-O       &GG 13   & SP  & 18303  & B$>$ 16.36  \\ 
\hline  
\end{tabular} 
\end{tiny}
\end{table*}
%%%%%%%%%%%%%%%%%%%%%%%%%%%%%%%%%%%%%%%%%%%%%%%%%%%%%%%%%%%%%%%%%%%%%%%%%%%%%%%%%%%%%%%%%%%%%%%%%%%

%%%%%%%%%%%%%%%%%%%%%%%%%%%%%%%%%%%%%%%%%%%%%%%%%%%%%%%%%%%%%%%%%%%%%%%%%%%%%%%%%%%%%%%%%%%%%%%%%%%
\begin{table*}
\caption{As in Table\ref{photometry_V2492:tab}, for V350 Cep. %In the last column, magnitudes are provided with the estimated error
\label{photometry_V350:tab}} 
\begin{tiny}
\begin{tabular}{c|ccr|llcr|l}
\hline
\hline
\multicolumn{1}{c}{HJD}&
\multicolumn{1}{c}{date}&
\multicolumn{1}{c}{UT}&
\multicolumn{1}{c}{expt}&
\multicolumn{1}{c}{emulsion}&
\multicolumn{1}{c}{filter}&
\multicolumn{1}{c}{tel}&
\multicolumn{1}{c}{plate}&
\multicolumn{1}{c}{mag}\\
\hline
  2441183.55876 & 1971-08-19    & 25:23 & 20    & 103a-O                & ....  & SG      & 04613&  B$>$19.33     \\ 
  2441183.58515 & 1971-08-19    & 26:01 & 20    & I-N sen               & RG 5  & SG      & 04614&  I$>$15.62     \\ 
  2441207.40035 & 1971-09-12    & 21:34 & 20    & 103a-O                & GG 13 & SG      & 04619&  B$=$18.69     \\ 
  2441207.45521 & 1971-09-12    & 22:53 & 30    & I-N sen               & RG 5  & SG      & 04621&  I$>$15.62     \\ 
  2441248.39933 & 1971-10-23    & 21:32 & 20    & 103a-O                & GG 13 & SG      & 04827&  B$>$19.33     \\ 
  2441248.42086 & 1971-10-23    & 22:03 & 30    & I-N sen               & RG 5  & SG      & 04828&  I$>$14.89     \\ 
  2441279.28642 & 1971-11-23    & 18:50 & 30    & I-N sen               & RG 5  & SG      & 04967&  I$=$15.18     \\ 
  2441279.35239 & 1971-11-23    & 20:25 & 20    & 103a-O                & GG 13 & SG      & 04970&  B$=$18.91     \\ 
  2441293.26107 & 1971-12-07    & 18:14 & 30    & I-N sen               & RG 5  & SG      & 05013&  I$=$15.18     \\ 
  2441293.28121 & 1971-12-07    & 18:43 & 20    & Ia-O                  & GG 13 & SG    & 05014&  B$>$17.21       \\ 
  2441298.30883 & 1971-12-12    & 19:23 & 20    & Ia-O                  & GG 13 & SG      & 05068&  B$>$17.21     \\ 
  2441298.32897 & 1971-12-12    & 19:52 & 30    & I-N sen               & RG 5  & SG      & 05069&  I$>$15.62     \\ 
  2441303.26285 & 1971-12-17    & 18:17 & 30    & I-N sen               & RG 5  & SG      & 05117&  I$=$15.26     \\ 
  2441303.28784 & 1971-12-17    & 18:53 & 20    & 103a-O                & GG 13 & SG      & 05118&  B$=$19.33     \\ 
  2441601.46877 & 1972-10-10    & 23:12 & 20    & 103a-O                & GG 13 & SG      & 05665&  B$>$17.21     \\ 
  2441601.49169 & 1972-10-10    & 23:45 & 30    & I-N sen               & RG 5  & SG      & 05666&  I$=$15.47     \\ 
  2441624.30411 & 1972-11-02    & 19:15 & 20    & 103a-O                & GG 13 & SG      & 05746&  B$>$17.21     \\ 
  2441624.32495 & 1972-11-02    & 19:45 & 30    & I-N sen               & RG 5  & SG    & 05747&  I$>$15.62       \\ 
  2441628.29157 & 1972-11-06    & 18:57 & 20    & 103a-O                & GG 13 & SG      & 05806&  B$=$19.12     \\ 
  2441628.31240 & 1972-11-06    & 19:27 & 30    & I-N sen               & RG 5  & SG      & 05807&  I$>$14.89     \\ 
  2441668.22049 & 1972-12-16    & 17:16 & 25    & I-N sen               & RG 5  & SG      & 05955&  I$=$15.55     \\ 
  2441665.43030 & 1972-12-13    & 22:18 & 05    & I-N sen               & RG 5  & SG      & 06003&  I$=$15.33     \\ 
  2441674.21960 & 1972-12-22    & 17:15 & 30    & I-N sen               & RG 5  & SG      & 06015&  I$>$14.45     \\ 
  2441674.23974 & 1972-12-22    & 17:44 & 20    & 103a-D                & GG 14 & SG      & 06016&  V$=$18.24     \\ 
  2441977.49586 & 1973-10-21    & 23:51 & 20    & 103a-O                & GG 13 & SG      & 06739&  B$>$17.21     \\ 
  2441977.51461 & 1973-10-21    & 24:18 & 30    & I-N sen               & RG 5  & SG      & 06740&  I$>$14.89     \\ 
  2441983.33958 & 1973-10-27    & 20:06 & 30    & I-N sen               & RG 5  & SG      & 06800&  I$=$15.55     \\ 
  2441983.36181 & 1973-10-27    & 20:38 & 20    & 103a-O                & GG 13 & SG      & 06801&  B$>$17.21     \\ 
  2441988.42565 & 1973-11-01    & 22:10 & 30    & I-N sen               & RG 5  & SG    & 06854&  I$>$14.89       \\ 
  2441988.44996 & 1973-11-01    & 22:45 & 20    & 103a-O                & GG 13 & SG    & 06855&  B$>$17.21       \\ 
  2442029.21438 & 1973-12-12    & 17:07 & 30    & 103a-D                & GG 14 & SG      & 06922&  V$=$17.72     \\ 
  2442034.26005 & 1973-12-17    & 18:13 & 20    & 103a-D                & GG 14 & SG      & 06945&  V$=$17.86     \\ 
  2442034.28088 & 1973-12-17    & 18:43 & 30    & I-N sen               & RG 5  & SG      & 06946&  I$=$15.04     \\ 
  2443074.36738 & 1976-10-22    & 20:46 & 30    & 103a-O                & GG 13 & SG      & 08724&  B$=$16.81     \\ 
  2443074.42225 & 1976-10-22    & 22:05 & 30    & I-N sen               & RG 5  & SG      & 08727&  I$=$14.25     \\ 
  2443135.31196 & 1976-12-22    & 19:28 & 30    & 103a-O                & GG 13 & SG      & 08818&  B$=$16.35     \\ 
  2443135.33696 & 1976-12-22    & 20:04 & 30    & I-N sen               & RG 5  & SG      & 08819&  I$=$13.82     \\ 
  2443421.44929 & 1977-10-04    & 22:44 & 30    & I-N sen               & RG 5  & SG      & 09235&  I$=$14.54     \\ 
  2443430.50003 & 1977-10-13    & 23:57 & 20    & 103a-O                & GG 13 & SG      & 09252&  B$=$16.02     \\ 
  2443430.52364 & 1977-10-13    & 24:31 & 30    & I-N sen               & RG 5  & SG      & 09253&  I$=$14.01     \\ 
  2443456.49987 & 1977-11-08    & 23:57 & 30    & I-N sen               & RG 5  & SG      & 09292&  I$=$14.37     \\ 
  2443456.52279 & 1977-11-08    & 24:30 & 20    & 103a-O                & GG 13 & SG      & 09293&  B$=$17.42     \\ 
  2443482.25558 & 1977-12-04    & 18:06 & 30    & 103a-O                & GG 13 & SG      & 09352&  B$=$16.68     \\ 
  2443482.28058 & 1977-12-04    & 18:42 & 30    & I-N sen               & RG 5  & SG      & 09353&  I$=$13.93     \\ 
  2443816.39439 & 1978-11-03    & 21:25 & 30    & 103a-D                & GG 14 & SG      & 09723&  V$=$16.02     \\ 
  2443837.31348 & 1978-11-24    & 19:29 & 20    & 103a-O                & GG 13 & SG      & 09793&  B$=$16.89     \\ 
  2443837.33500 & 1978-11-24    & 20:00 & 30    & 103a-E                & RG 1  & SG      & 09794&  R$=$15.09     \\ 
  2443842.35919 & 1978-11-29    & 20:35 & 30    & 103a-E                & RG 1  & SG      & 09804&  R$=$15.02     \\ 
  2443845.35008 & 1978-12-02    & 20:22 & 30    & 103a-E                & RG 1  & SG      & 09825&  R$=$15.30     \\ 
  2443845.37231 & 1978-12-02    & 20:54 & 20    & 103a-O                & GG 13 & SG      & 09826&  B$=$18.06     \\   
  2444116.51532 & 1979-08-30    & 24:20 & 20    & 103a-O                & GG 13 & SG      & 10164&  B$=$17.63     \\ 
  2444116.53755 & 1979-08-30    & 24:52 & 30    & I-N sen               & RG 5  & SG      & 10165&  I$=$14.37     \\ 
  2444146.36243 & 1979-09-29    & 20:39 & 30    & I-N sen               & RG 5  & SG    & 10197&  I$=$14.08       \\ 
  2444146.39090 & 1979-09-29    & 21:20 & 20    & 103a-O                & GG 13 & SG      & 10198&  B$=$16.57     \\ 
  2444514.28954 & 1980-10-01    & 18:54 & 30    & I-N sen               & RG 5  & SG      & 10629&  I$>$14.45     \\ 
  2444514.33746 & 1980-10-01    & 20:03 & 30    & 103a-O                & GG 13 & SG      & 10630&  B$=$16.35     \\ 
  2444551.49850 & 1980-11-07    & 23:55 & 20    & 103a-O                & GG 13 & SG      & 10662&  B$=$16.68     \\ 
  2444551.52072 & 1980-11-07    & 24:27 & 30    & I-N sen               & RG 5  & SG    & 10663&  I$=$14.08       \\ 
  2444819.49780 & 1981-08-02    & 23:56 & 30    & 103a-O                & GG 13 & SG      & 11056&  B$=$16.68     \\ 
  2444853.32451 & 1981-09-05    & 19:45 & 30    & I-N sen               & RG 5  & SG    & 11089&  I$=$13.93       \\ 
  2444853.35090 & 1981-09-05    & 20:23 & 30    & 103a-O                & GG 13 & SG      & 11090&  B$=$16.46     \\ 
  2444913.36104 & 1981-11-04    & 20:37 & 30    & I-N sen               & RG 5  & SG    & 11168&  I$=$14.20       \\ 
  2444926.33862 & 1981-11-17    & 20:05 & 30    & I-N sen               & RG 5  & SG    & 11191&  I$=$14.16       \\ 
  2444926.36292 & 1981-11-17    & 20:40 & 30    & 103a-O                & GG 13 & SG      & 11192&  B$=$16.68     \\ 
  2445591.38996 & 1983-09-13    & 21:19 & 30    & I-N sen               & RG 5  & SG      & 12208&  I$=$13.93     \\ 
  2446293.45032 & 1985-08-15    & 22:47 & 30    & IIa-O                 & GG 13 & SG      & 12930&  B$=$16.68     \\ 
  2446295.46427 & 1985-08-17    & 23:07 & 30    & I-N sen               & RG 5  & SG    & 12939&  I$=$13.93       \\ 
  2446297.44280 & 1985-08-19    & 22:36 & 30    & I-N sen               & RG 5  & SG      & 12948&  I$=$14.16     \\ 
\hline
\end{tabular}
\end{tiny}
\end{table*}
%%%%%%%%%%%%%%%%%%%%%%%%%%%%%%%%%%%%%%%%%%%%%%%%%%%%%%%%%%%%%%%%%%%%%%%%%%%%%%%%%%%%%%%%%%%%%%%%%%%

%%%%%%%%%%%%%%%%%%%%%%%%%%%%%%%%%%%%%%%%%%%%%%%%%%%%%%%%%%%%%%%%%%%%%%%%%%%%%%%%%%%%%%%%%%%%%%%%%%%
\begin{table*} 
\caption{As in Table\ref{photometry_V2492:tab}, for ASASSN-15qi.
% In the last column, magnitudes are provided with the estimated error  
\label{photometry_ASASSN:tab}} 
\begin{tiny}
\begin{tabular}{c|ccr|llcr|l}
\hline
\hline
\multicolumn{1}{c}{HJD}&
\multicolumn{1}{c}{date}&
\multicolumn{1}{c}{UT}&
\multicolumn{1}{c}{expt}&
\multicolumn{1}{c}{emulsion}&
\multicolumn{1}{c}{filter}&
\multicolumn{1}{c}{tel}&
\multicolumn{1}{c}{plate}&
\multicolumn{1}{c}{mag}\\
\hline
2440529.26267 & 1969-11-03      & 18:14  & 30   & 103 a-O       & GG 13 & SG & 02750  & B$=$17.90 \\
2440529.28697 & 1969-11-03      & 18:49  & 30   & I-N sen       & RG 5  & SG & 02751  & I$=$15.32 \\
2440836.43372 & 1970-09-06      & 22:21  & 30   & 103 a-O       & GG 13 & SG & 03617  & B$=$18.57 \\
2440836.46011 & 1970-09-06      & 22:59  & 30   & I-N sen       & RG 5  & SG & 03618  & I$=$14.13 \\
2440882.44129 & 1970-10-22      & 22:31  & 30   & 103 a-O       & GG 13 & SG & 03757  & B$=$17.64 \\
2440882.46560 & 1970-10-22      & 23:06  & 30   & I-N sen       & RG 5  & SG & 03758  & I$=$14.35 \\
2440885.41835 & 1970-10-25      & 21:58  & 30   & 103 a-O       & GG 13 & SG & 03781  & B$>$17.73 \\
2440888.48568 & 1970-10-28      & 23:35  & 30   & I-N sen       & RG 5  & SG & 03826  & I$=$15.23 \\
2440888.51207 & 1970-10-28      & 24:13  & 30   & 103 a-O       & GG 13 & SG & 03827  & B$=$17.56 \\
2440895.43001 & 1970-11-04      & 22:15  & 40   & 103 a-E       & RG 1  & SG & 03861  & R$=$14.92 \\
2440895.45779 & 1970-11-04      & 22:55  & 30   & 103 a-O       & GG 13 & SG & 03862  & B$=$17.56 \\
2440918.23491 & 1970-11-27      & 17:35  & 30   & I-N sen       & RG 5  & SG & 03951  & I$=$15.15 \\
2440918.25922 & 1970-11-27      & 18:10  & 30   & 103 a-O       & GG 13 & SG & 03952  & B$=$18.57 \\
2440924.31731 & 1970-12-03      & 19:34  & 31   & I-N sen       & RG 5  & SG & 04019  & I$=$14.86 \\
2440924.34231 & 1970-12-03      & 20:10  & 30   & 103 a-O       & GG 13 & SG & 04020  & B$=$18.32 \\
2440941.25820 & 1970-12-20      & 18:10  & 40   & II a-O        & GG 13 & SG & 04084  & B$=$18.32 \\
2440941.28597 & 1970-12-20      & 18:50  & 30   & I-N sen       & RG 5  & SG & 04085  & I$=$14.86 \\
2440970.23026 & 1971-01-18      & 17:32  & 30   & I-N sen       & RG 5  & SG & 04149  & I$=$14.98 \\
2441179.47520 & 1971-08-15      & 23:22  & 30   & 103 a-O       & GG 13 & SG & 04574  & B$>$17.73 \\
2441179.50367 & 1971-08-15      & 24:03  & 40   & 103 a-E       & RG 1  & SG & 04575  & R$=$15.66 \\
2441179.53284 & 1971-08-15      & 24:45  & 30   & I-N sen       & RG 5  & SG & 04576  & I$=$15.20 \\
2441215.47854 & 1971-09-20      & 23:25  & 30   & 0a-O          & GG 13 & SG & 04660  & B$=$18.32 \\
2441215.50910 & 1971-09-20      & 24:09  & 40   & I-N sen       & RG 5  & SG & 04661  & I$=$15.20 \\
2441245.40311 & 1971-10-20      & 21:36  & 20   & 103 a-O       & GG 13 & SG & 04796  & B$=$17.90 \\
2441245.42325 & 1971-10-20      & 22:05  & 30   & I-N sen       & RG 5  & SG & 04797  & I$=$14.86 \\
2441272.39362 & 1971-11-16      & 21:23  & 30   & 103 a-O       & GG 13 & SG & 04934  & B$=$18.49 \\
2441272.42556 & 1971-11-16      & 22:09  & 40   & I-N sen       & RG 5  & SG & 04935  & I$=$15.19 \\
2441291.23044 & 1971-12-05      & 17:29  & 25   & 103 a-O       & GG 13 & SG & 04998  & B$=$18.57 \\
2441292.21373 & 1971-12-06      & 17:05  & 35   & I-N sen       & RG 5  & SG & 05004  & I$=$15.11 \\
2441292.23803 & 1971-12-06      & 17:40  & 25   & 103 a-O       & GG 13 & SG & 05005  & B$=$18.49 \\
2441297.24753 & 1971-12-11      & 17:54  & 40   & I-N sen       & RG 5  & SG & 05050  & I$=$14.98 \\
2441297.27600 & 1971-12-11      & 18:35  & 30   & Ia-O          & GG 13 & SG & 05051  & B$=$18.57 \\
2441300.20851 & 1971-12-14      & 16:58  & 40   & I-N sen       & RG 5  & SG & 05078  & I$=$14.98 \\
2441300.23628 & 1971-12-14      & 17:38  & 30   & Ia-O          & GG 13 & SG & 05079  & B$>$17.73 \\
2441570.44841 & 1972-09-09      & 22:42  & 20   & 103 a-O       & GG 13 & SG & 05598  & B$=$18.57 \\
2441570.47411 & 1972-09-09      & 23:19  & 30   & I-N sen       & RG 5  & SG & 05599  & I$=$14.98 \\
2441606.40035 & 1972-10-15      & 21:32  & 30   & I-N sen       & RG 5  & SG & 05682  & I$=$15.28 \\
2441606.42049 & 1972-10-15      & 22:01  & 20   & 103 a-O       & GG 13 & SG & 05683  & B$>$17.73 \\
2441622.39813 & 1972-10-31      & 21:29  & 20   & 103 a-O       & GG 13 & SG & 05724  & B$=$18.57 \\
2441622.42174 & 1972-10-31      & 22:03  & 30   & I-N sen       & RG 5  & SG & 05725  & I$=$15.03 \\
2441626.34181 & 1972-11-04      & 20:08  & 20   & 103 a-O       & GG 13 & SG & 05776  & B$=$17.54 \\
2441626.36750 & 1972-11-04      & 20:45  & 30   & I-N sen       & RG 5  & SG & 05777  & I$=$15.12 \\
2441649.22587 & 1972-11-27      & 17:22  & 30   & I-N sen       & RG 5  & SG & 05908  & I$=$14.92 \\
2441649.24601 & 1972-11-27      & 17:51  & 20   & 103 a-O       & GG 13 & SG & 05909  & B$=$18.57 \\
2441659.28309 & 1972-12-07      & 18:45  & 20   & 103 a-D       & GG 14 & SG & 05972  & V$=$16.01 \\
2441659.30254 & 1972-12-07      & 19:13  & 30   & I-N sen       & RG 5  & SG & 05973  & I$=$14.92 \\
2441665.45712 & 1972-12-13      & 22:56  & 30   & I-N sen       & RG 5  & SG & 06004  & I$=$14.92 \\
2441665.47725 & 1972-12-13      & 23:25  & 30   & 103 a-D       & GG 14 & SG & 06005  & V$=$15.85 \\
2441675.21358 & 1972-12-23      & 17:06  & 30   & I-N sen       & RG 5  & SG & 06018  & I$=$14.98 \\
2441675.23372 & 1972-12-23      & 17:35  & 20   & 103 a-D       & GG 14 & SG & 06019  & V$=$16.24 \\
2441958.54820 & 1973-10-02      & 25:05  & 30   & I-N sen       & RG 5  & SG & 06671  & I$>$15.04 \\
2441958.57111 & 1973-10-02      & 25:38  & 20   & 103 a-O       & GG 13 & SG & 06672  & B$=$18.57 \\
2441976.42464 & 1973-10-20      & 22:07  & 30   & 103 a-D       & GG 14 & SG & 06733  & V$=$16.43 \\
2441976.46491 & 1973-10-20      & 23:05  & 20   & 103 a-O       & GG 13 & SG & 06734  & B$>$16.57 \\
2441980.47391 & 1973-10-24      & 23:18  & 30   & I-N sen       & RG 5  & SG & 06767  & I$=$14.80 \\
2441980.49683 & 1973-10-24      & 23:51  & 20   & 103 a-O       & GG 13 & SG & 06768  & B$>$17.73 \\
2441983.38291 & 1973-10-27      & 21:07  & 20   & 103 a-O       & GG 13 & SG & 06802  & B$=$18.57 \\
2441983.40166 & 1973-10-27      & 21:34  & 20   & 103 a-D       & GG 14 & SG & 06803  & V$=$16.09 \\
2441987.46410 & 1973-10-31      & 23:04  & 20   & 103 a-O       & GG 13 & SG & 06850  & B$>$18.57 \\
2441987.48563 & 1973-10-31      & 23:35  & 30   & I-N sen       & RG 5  & SG & 06851  & I$>$15.86 \\
2442002.31516 & 1973-11-15      & 19:30  & 20   & 103 a-D       & GG 14 & SG & 06879  & V$=$16.41 \\
2442017.39450 & 1973-11-30      & 21:25  & 30   & 103 a-D       & GG 14 & SG & 06899  & V$=$15.95 \\
2442017.42228 & 1973-11-30      & 22:05  & 30   & I-N sen       & RG 5  & SG & 06900  & I$>$15.04 \\
2442029.24260 & 1973-12-12      & 17:47  & 20   & 103 a-D       & GG 14 & SG & 06923  & V$=$16.67 \\
2442029.26760 & 1973-12-12      & 18:23  & 30   & I-N sen       & RG 5  & SG & 06924  & I$>$13.99 \\
2442034.30694 & 1973-12-17      & 19:20  & 30   & I-N sen       & RG 5  & SG & 06947  & I$=$15.41 \\
2442034.32777 & 1973-12-17      & 19:50  & 20   & 103 a-D       & GG 14 & SG & 06948  & V$=$16.51 \\
2442689.36765 & 1975-10-03      & 20:45  & 30   & 103 a-O       & GG 13 & SG & 08102  & B$=$17.90 \\
2442689.38917 & 1975-10-03      & 21:16  & 30   & I-N sen       & RG 5  & SG & 08103  & I$=$14.39 \\
2442712.38848 & 1975-10-26      & 21:15  & 30   & 103 a-O       & GG 13 & SG & 08151  & B$=$18.57 \\
2442712.41626 & 1975-10-26      & 21:55  & 30   & I-N sen       & RG 5  & SG & 08152  & I$=$14.92 \\
2442746.31470 & 1975-11-29      & 19:30  & 20   & 103 a-O       & GG 13 & SG & 08206  & B$=$18.57 \\
\hline
\end{tabular}
\end{tiny}
%\tablefoot{
%\tablefoottext{a}{A},\tablefoottext{b}{B},\tablefoottext{c}{C},\tablefoottext{d}{D}\\
%}
\end{table*}  
\setcounter{table}{2}
\begin{table*}
\caption{Continued.} 
\begin{tiny}
\begin{tabular}{c|ccr|llcr|l}
\hline
\hline
2442746.33693 & 1975-11-29      & 20:02  & 30   & I-N sen       & RG 5  & SG & 08207  & I$=$14.92 \\
2443039.46808 & 1976-09-17      & 23:10  & 20   & 103 a-O       & GG 13 & SG & 08684  & B$=$18.07 \\
2443039.48961 & 1976-09-17      & 23:41  & 30   & I-N sen       & RG 5  & SG & 08685  & I$=$15.56 \\
2443074.31698 & 1976-10-22      & 19:32  & 30   & I-N sen       & RG 5  & SG & 08722  & I$=$14.86 \\
2443074.34129 & 1976-10-22      & 20:07  & 30   & 103 a-O       & GG 13 & SG & 08723  & B$=$18.49 \\
2443127.39735 & 1976-12-14      & 21:30  & 30   & 103 a-O       & GG 13 & SG & 08797  & B$>$17.73 \\
2443127.42165 & 1976-12-14      & 22:05  & 30   & I-N sen       & RG 5  & SG & 08798  & I$=$14.80 \\
2443392.47259 & 1977-09-05      & 23:17  & 30   & I-N sen   & RG 5      & SG & 09166  & I$=$15.12 \\
2443392.49481 & 1977-09-05      & 23:49  & 20   & 103 a-O   & GG 13     & SG & 09167  & B$>$17.73 \\
2443398.46027 & 1977-09-11      & 22:59  & 30   & I-N sen       & RG 5  & SG & 09199  & I$=$14.92 \\
2443398.48527 & 1977-09-11      & 23:35  & 20   & 103 a-O       & GG 13 & SG & 09200  & B$>$17.73 \\
2443405.43198 & 1977-09-18      & 22:18  & 20   & 103 a-O       & GG 13 & SG & 09213  & B$=$17.54 \\
2443405.46254 & 1977-09-18      & 23:02  & 30   & I-N sen       & RG 5  & SG & 09214  & I$=$15.38 \\
2443420.40376 & 1977-10-03      & 21:37  & 30   & I-N sen       & RG 5  & SG & 09224  & I$=$14.92 \\
2443452.27862 & 1977-11-04      & 18:37  & 30   & I-N sen       & RG 5  & SG & 09273  & I$=$14.26 \\
2443464.35194 & 1977-11-16      & 20:23  & 30   & I-N sen       & RG 5  & SG & 09324  & I$=$14.92 \\
2443480.35415 & 1977-12-02      & 20:27  & 30   & 103 a-O       & GG 13 & SG & 09347  & B$>$16.88 \\
2443480.43262 & 1977-12-02      & 22:20  & 30   & I-N sen       & RG 5  & SG & 09350  & I$=$15.29 \\
2443492.30361 & 1977-12-14      & 19:15  & 30   & I-N sen       & RG 5  & SG & 09372  & I$>$15.04 \\
2443492.38139 & 1977-12-14      & 21:07  & 30   & 103 a-O       & GG 13 & SG & 09375  & B$>$17.73 \\
2443789.40935 & 1978-10-07      & 21:45  & 30   & 103 a-D       & GG 14 & SG & 09652  & V$=$16.38 \\
2443789.43366 & 1978-10-07      & 22:20  & 30   & I-N sen       & RG 5  & SG & 09653  & I$=$15.80 \\
2444116.38069 & 1979-08-30      & 21:05  & 30   & I-N sen       & RG 5  & SG & 10162  & I$=$15.19 \\
2444116.49528 & 1979-08-30      & 23:50  & 20   & 103 a-O       & GG 13 & SG & 10163  & B$>$17.73 \\
2444143.44325 & 1979-09-26      & 22:34  & 30   & I-N sen       & RG 5  & SG & 10186  & I$=$14.98 \\
2444143.47172 & 1979-09-26      & 23:15  & 20   & 103 a-O       & GG 13 & SG & 10187  & B$>$17.73 \\
2444172.45030 & 1979-10-25      & 22:44  & 20   & 103 a-O       & GG 13 & SG & 10249  & B$>$17.73 \\
2444172.47182 & 1979-10-25      & 23:15  & 30   & I-N sen       & RG 5  & SG & 10250  & I$=$14.98 \\
2444552.47852 & 1980-11-08      & 23:25  & 20   & 103 a-D       & GG 14 & SG & 10668  & V$=$15.62 \\
2444557.42771 & 1980-11-13      & 22:12  & 20   & 103 a-O       & GG 13 & SG & 10672  & B$>$16.75 \\
2444557.45479 & 1980-11-13      & 22:51  & 30   & I-N sen       & RG 5  & SG & 10673  & I$>$15.04 \\
2444854.40387 & 1981-09-06      & 21:38  & 30   & I-N sen       & RG 5  & SG & 11099  & I$=$15.19 \\
2444911.34879 & 1981-11-02      & 20:18  & 30   & I-N sen       & RG 5  & SG & 11152  & I$=$15.26 \\
2444911.37726 & 1981-11-02      & 20:59  & 30   & 103 a-O       & GG 13 & SG & 11153  & B$=$17.85 \\
2444933.41070 & 1981-11-24      & 21:48  & 30   & 103 a-O       & GG 13 & SG & 11241  & B$>$17.73 \\
2444933.44195 & 1981-11-24      & 22:33  & 30   & I-N sen       & RG 5  & SG & 11242  & I$>$15.04 \\
2445264.32394 & 1982-10-21      & 19:42  & 25   & I-N sen       & RG 5  & SG & 11723  & I$=$14.86 \\
2445264.35310 & 1982-10-21      & 20:24  & 30   & 103 a-D       & GG 14 & SG & 11724  & V$=$16.38 \\
2445290.37764 & 1982-11-16      & 21:00  & 30   & 103 a-O       & GG 13 & SG & 11740  & B$=$17.98 \\
2445591.41726 & 1983-09-13      & 21:57  & 30   & I-N sen       & RG 5  & SG & 12209  & I$=$15.20 \\
2445591.44296 & 1983-09-13      & 22:34  & 30   & 103 a-D       & GG 14 & SG & 12210  & V$>$16.63 \\
2445611.46209 & 1983-10-03      & 23:01  & 30   & 103 a-O       & GG 13 & SG & 12228  & B$=$17.67 \\
2449223.38669 & 1993-08-23      & 21:14  & 30   & 103 a-E       & RG 1  & SG & 15636  & R$>$15.05 \\
\hline
\end{tabular}
\end{tiny}
\end{table*}
%%%%%%%%%%%%%%%%%%%%%%%%%%%%%%%%%%%%%%%%%%%%%%%%%%%%%%%%%%%%%%%%%%%%%%%%%%%%%%%%%%%%%%%%%%%%%%%%%%%

%%%%%%%%%%%%%%   TABLE 4 - STATISTICS %%%%%%%%%%%%%%%%%%%%%%%%%%%%%%%%%

\begin{table*}
\caption{Ranges of photometric variability for ASASSN-15qi. For each band, we list the number of observations not considering the upper limits (Column 3), the median, which basically indicates the magnitude in quiescence (Column 4), the standard deviation data point distribution (Column 5), and the magnitude corresponding to the peak brightness (Column 6). \label{ranges:tab}} 
\begin{center} 
\medskip
{
%\scriptsize
\begin{tabular}{l|c|c|ccc}
\hline
Source          &  Band     & N$_{obs}$ & Median & $\sigma$ &Peak  \\ 
                &           &           &          \multicolumn{3}{c}{(mag)    }     \\
\hline
                &             &           &               &         &                   \\
ASASSN-15qi     &   $B$       &  29       &  18.32        &  0.40   &  17.54            \\
                &   $V$       &  12       &  16.38        &  0.32   &  15.62            \\ 
                &   $R$       &   3       &  15.05        &  0.39   &  14.92            \\  
                &   $I$       &  45       &  14.98        &  0.31   &  14.13            \\         
\hline
\end{tabular}}
\end{center}
\end{table*}    
%%%%%%%%%%%%%%%%%%%%%%%%%%%%%%%%%%%%%%%%%%%%%%%%%%%%%%%%%%%%%%%%%%%%%%%

\begin{figure*}
\centering
%\vspace*{-7cm}
\includegraphics[width=13cm, trim = 0cm 0cm 0cm 9cm, clip]{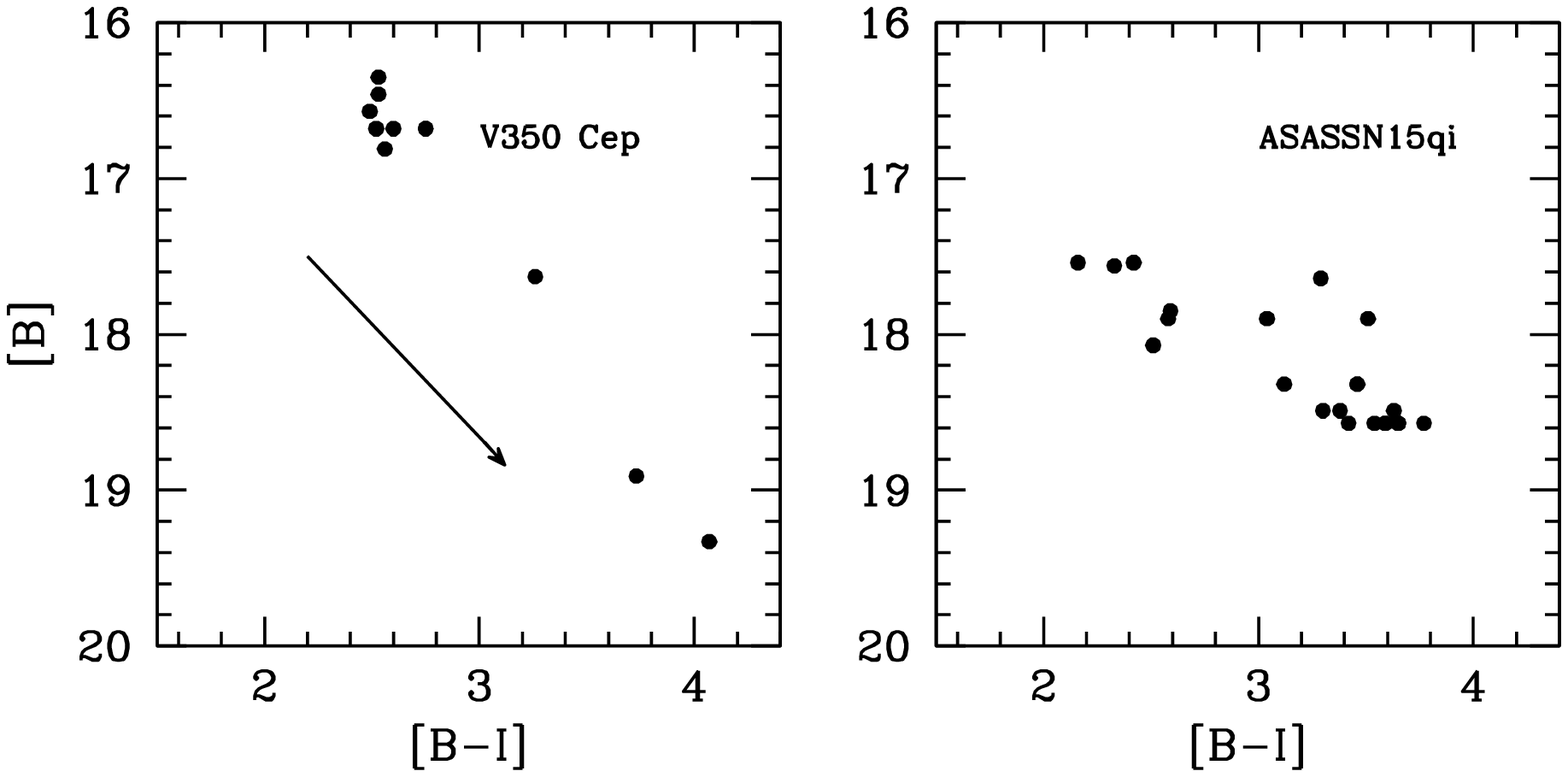}
\caption{\label{B_BI_plot:fig} $B$ vs. [$B-I$] colour-magnitude plot of V350 Cep (left) and ASASSN-15qi (right). 
In the lower-left corner the arrow indicates an extinction of A$_V$ = 1 mag, according to the law by Rieke \& Lebofsky (1985).}
\end{figure*}
 
\section{Analysis and discussion}

\subsection{V2492 Cyg}

During our monitoring period V2492 Cyg remained always undetected at our sensitivity, therefore the light curve 
depicted in Fig.~\ref{V2492_light:fig} displays only upper limit values at different levels, as explained in Sect. 3. As a consequence, no fading or outbursting event can be detected, nevertheless some useful information can be derived. As mentioned above, after
its discovery in 2010, V2492 underwent a long-lasting period of strong activity with intermittent burst and fading events (see Hillenbrand et al. 2013 and AAVSO \footnote{American Association of Variable Star Observers \\(https://www.aavso.org)} data) and 
reached its maximum recorded brightness in 2017 (Giannini et al. 2017). During most of this period, the source was 
sampled with an almost daily cadence and, for long time intervals, remained brighter than the following values: $B$<18, $V$<16, $R$<15, and $I$<14 mag. 
In comparison, our plate measurements are largely undersampled, presenting long periods (up to a decade) without any data. 
In any case (not considering the $I$ band, which is practically uncovered, and the $B$ band, which presents no significant upper limits), our $V$ and $R$-band upper 
limits tentatively suggest that duringa period of about 30 yr from 1958 to 1987 an activity similar (both in duration 
and in brightness) to that more recently (2010-2017) monitored, did not occur.
%By analyzing the recent data, Hillenbrand et al. suggested that the activity of V2492 Cyg might have a period of about 220 days. 
%To give a quantitative support to our interpretation we have repeatedly superimposed (as a vertically dashed plot in Figure~\ref{V2492_light:fig}) the recent activity plot onto our upper limits one, going back in time hypothesizing that such an activity obeys that periodicity. As can be seen (central panels of Figure~\ref{V2492_light:fig}), upper limits seem to exclude that a similar activity occurred in the past, as well. 
%REFEREE: Subsequent observations of the star showed no periodicity of around two hundred days, or similar
Indeed, 
%only an activity at a definitely fainter level (by more than 3 mag) could be still compatible with our results. 
%even considering the recent activity not in terms of periodicity, 
we note that for a significant amount of time the source is brighter than the level indicated by our upper limits, thus suggesting that, in the past, the activity of V2492 Cyg was not as strong as it is now. The above scenario, if correctly described, means that an enhanced brightness variability 
%(especially if periodic) 
could be an infrequent feature of V2492 Cyg. Such circumstances favour an accretion- more than an extinction-driven origin for the bursts. Indeed, the former is expected to occur with a long and irregular cadence related to the viscous motion of the matter toward the inner edge of the disk, while the latter should occur more frequently and regularly, according to the orbital motion of the obscuring matter along the line of sight.

\subsection{V350 Cep} 

As depicted in Fig.~\ref{V350_light:fig}, our 15-year coverage (1971-1985) presents a significant sequence of data  in the $B$ and $I$ bands only. These data substantially confirm the behaviour of V350 Cep illustrated in Sect.1: namely a burst occurred in the middle of the 1970s, after which V350 became well detectable, at a roughly constant level of brightness. To our knowledge, the available
literature does not point out any sign of a peculiar activity until the sudden fading and quick restoring of its maximum brightness,
occurring in the $B$ band during 2016 (Semkov et al 2017). Our data indicate this has not been a unique event. 
Indeed, between 1977 and 1980, in the same band, there is evidence of intermittent luminosity variations of $\Delta$B~$\sim$~1-1.5 mag 
on a time-scale of months. A lesser variation is barely recognisable also in the I band. The existence of (at least) two fading episodes
in few decades makes it difficult to classify V350 Cep as a pure FUor star, although its relevant brightening occurred in the 1970s. On the other hand, albeit increasing the monitoring coverage, the lack of significant outbursts remains confirmed, and therefore, also the EXor hypothesis does not seem viable. A longer and continuous multi-band monitoring could provide a more certain classification, but, at the present stage, we note that the presence of an orbiting structure along the line of sight (UXor hypothesis) has a role in determining its light curve. The strict similarity between the two fading episodes highlighted by our observations corroborates such a conclusion. 

\subsection{ASASSN-15qi}  

Although much more frequently sampled and for a much longer period, our data substantially agree with the archival photometry given by Herczeg et al. (2016) (see their Table 1). Our monitoring refers to a long-lasting (1970-1983) quiescence period during which brightness fluctuations (up to 1 mag) in the B, V, and I bands occurred. Therefore, ASASSN-15qi behaved in the past as a moderately active young source similar to the classical T Tauri stars, albeit more massive and luminous. It should be noted that a very fast event (lasting about 10 days) analogous to that recently pointed out by Maehara et al. (2015) (see Sect.1) would, theoretically, never have been detected by our monitoring, whose sampling is not frequent enough. As a consequence, we can only confirm the doubts already expressed by Herczeg et al. (2016) about the classification of this object. However, we can provide an argument related to the colour-magnitude plot (see following section 4.4) not supporting a pure UXor classification.

\subsection{Colour-magnitude diagrams} 
 
Given the available data we can only build the colour-magnitude plots $B$ versus [$B-I$]  for the sources V350 Cep and ASASSN-15qi, using
only magnitudes obtained within 1 day in $B$ and $I$ bands. They are given in Fig.~\ref{B_BI_plot:fig} and essentially support the
above considerations. In fact, while data points of V350 Cep (left panel) are well aligned along the extinction vector, ASASSN-15qi (right panel) comparatively shows a more dispersed distribution, typical of sources whose fluctuations cannot be reconciled with a pure extinction
origin.

\section{Concluding remarks}

Archival plate analysis is a tool well suited to searching the past history of young variables identified as eruptive stars, and hence to improve their classification.
We investigated the Asiago Schmidt plate collection for observations of the selected fields where the three eruptive sources V2492 Cyg, V350 Cep, and ASASSN-15qi are located. Observations of these regions were repeatedly carried out at Asiago over various time periods from 12 to 30 years. The analysis of Harvard plates of the same sources rules out the occurrence of large outbursts in the past.
We provide one of the best-sampled photometric datasets ever obtained of the past history of the three targets. In particular, V2492 historical upper limits do not seem compatible with the level of brightness of the present activity. Hence, this latter could have appeared only recently, possibly dominated by accretion phenomena more than by repetitive obscuration. For V350, our monitoring has pointed out
that sudden fading followed by rapid restoring of the previous brightness is not a unique and isolated event, and, as such, it may be attributable to an orbiting structure along the line of sight.  During the monitored quiescence, ASASSN-15qi presents a level of moderate variability (0.5-1 mag) that is comparable with that of classical T Tauri stars. We are not able to fill in any gaps existing in the
available literature about its nature. We can only say that a pure extinction origin does not seem to be the only mechanism responsible for the observed
fluctuations.

%~ thanks to people from Harvard College Observatory and AAVSO ?
\begin{acknowledgements}
We acknowledge with thanks the variable star observations from the AAVSO International Database contributed by observers worldwide and used in this research. \\

We thank the ASAS-SN project that is supported by the Gordon and Betty Moore Foundation through grant GBMF5490 to the Ohio State University and NSF grant AST-1515927.\\

We are thankful to the people of the Harvard College Observatory for their kind support.\\

R.J.S. thanks the INAF Astronomical Observatory of Padova for the hospitality during the stay in Asiago and for permission to use the 
historical plate archive of the Asiago Observatory. This work was supported in part by the Croatian Science Foundation under the project 
6212 Solar and Stellar Variability and by the University of Rijeka under the project number 13.12.1.3.03.
\end{acknowledgements}

\end{document}